\documentclass[aps,prl,preprint,groupedaddress]{revtex4-1}
\usepackage[dvipdfmx]{graphicx}% Include figure files
\usepackage{dcolumn}% Align table columns on decimal point
\usepackage{bm}% bold math

\begin{document}

% Use the \preprint command to place your local institutional report
% number in the upper righthand corner of the title page in preprint mode.
% Multiple \preprint commands are allowed.
% Use the 'preprintnumbers' class option to override journal defaults
% to display numbers if necessary
%\preprint{}

%Title of paper
\title{Softening Mechanism of Coherent Phonons in Antimony 
        \\ Under High Density Photoexcitation}

% repeat the \author .. \affiliation  etc. as needed
% \email, \thanks, \homepage, \altaffiliation all apply to the current
% author. Explanatory text should go in the []'s, actual e-mail
% address or url should go in the {}'s for \email and \homepage.
% Please use the appropriate macro foreach each type of information

% \affiliation command applies to all authors since the last
% \affiliation command. The \affiliation command should follow the
% other information
% \affiliation can be followed by \email, \homepage, \thanks as well.
\author{H. Kumagai} % \email{kuma@phys.sci.hokudai.ac.jp}
\author{I. Matsubara}
\author{J. Nakahara}
\author{T. Mishina} \email{mis@phys.sci.hokudai.ac.jp}

%\email[]{Your e-mail address}
%\homepage[]{Your web page}
%\thanks{}
%\altaffiliation{}
\affiliation{Department of Physics, Faculty of Science, Hokkaido University, 
             Sapporo 060-0810, Japan}

%Collaboration name if desired (requires use of superscriptaddress
%option in \documentclass). \noaffiliation is required (may also be
%used with the \author command).
%\collaboration can be followed by \email, \homepage, \thanks as well.
%\collaboration{}
%\noaffiliation

\date{\today}

\begin{abstract}
 We have investigated the dynamical properties of coherent phonons
generated in antimony under high density photo excitation. Precise
measurements and extended analysis of coherent phonon oscillations provide
new insights into the process of impulsive softening. In the process of recovering
from instantaneous softening, phonon frequency shows an
unexpected temporal evolution involving an abrupt change and a slight overshoot.
Moreover, fluence dependence of the initial phonon frequency is strongly
correlated with the phonon frequency shift of the high-pressure Raman spectra.
These results clearly prove that the structural changes (lattice contraction
and expansion) induced by laser excitation governs the softening mechanism
of coherent phonons.
\end{abstract}

% insert suggested PACS numbers in braces on next line
% insert suggested keywords - APS authors don't need to do this
%\keywords{}

%\maketitle must follow title, authors, abstract, \pacs, and \keywords
\maketitle

% body of paper here - Use proper section commands
% References sh

%\section{}
% Put \label in argument of \section for cross-referencing
%\section{\label{}}
%\subsection{}
%\subsubsection{}

%\section{Introduction}

 Photoexcitation of materials with intense femtosecond
laser pulses has led to a unique physical process in various materials 
on a femtosecond time scale. 
For example, phase transition \cite{Pashkin11},
laser ablation \cite{Carbone08, Otis91, Wenqian10}, 
laser melting \cite{Sokolowski-Tinten98, Wai-Lun08}, 
and band-gap renormalization \cite{Kleinman85, Nowak11}
have been intensively studied.
Softening of coherent phonons under high density excitation is one
of the most important and promising phenomena to investigate.
In addition to the continuous efforts to clarify the generation 
and detection mechanism of coherent phonons of 
various materials \cite{Cho90, Pfeifer92, Mishina00}, 
their softening mechanism under high density excitation 
has been intensively focused. 
Early observations of the softening of coherent phonons
in antimony and tellurium have been reported, and the mechanisms 
have been deduced as ionic screening 
by photoexcited carriers and purely electronic softening 
of the crystal lattice, respectively \cite{Cheng94, Hunsche95}. 
Hase {\it et al}. have proposed that the softening mechanism involves 
anharmonicity of the lattice potential
instead of electronic softening of the crystal lattice \cite{Hase02}.
The effect of photoexcited electron-hole plasma on the phonon dispersion 
relation of bismuth has been investigated using 
first principles density-functional perturbation theory,
and its influence on phonon softening also has been discussed \cite{Murray07}.
Recently, time-resolved X-ray diffraction experiments have revealed 
the atomic motion of bismuth under phonon softening conditions \cite{Fritz07}.
However, the existing experimental evidence is still insufficient to arrive 
at a final conclusion on the phonon softening mechanism. 

 In this study, we explore the precise measurement 
of coherent phonon oscillations in antimony under high density excitation 
and examine the details of time evolution of frequency shifts.
Time evolution exhibits an abrupt change and a slight overshoot,
which are hardly expected from a carrier-phonon interaction.
Moreover, fluence dependence of the initial frequency
is closely correlated with the high-pressure Raman experimental results.
The correlation is well explained by the theory based on the quantum motion of
nuclei under intense excitation of bonding electrons.
These results clearly prove that the structural changes induced
by laser excitation govern the softening mechanism of coherent phonons.

 The sample used in our experiment is a single crystal of antimony,
with its surface perpendicular to the trigonal axis.
Our experiments are performed in a reflection-type pump-probe system
at room temperature.
A commercial Ti:sapphire oscillator operating at repetition rates of 80 MHz
and pulse width of 100 fs is used as the excitation light source.
The pump and probe beams are orthogonally polarized to each other
and they are coaxially arranged using a polarization beam splitter.
The beams are focused on the sample to a diameter of about 3$ \mu $m
with a microscope objective lens having a focal length of 4.5 mm.
A high density excitation condition of maximum 20 mJ/cm$^{2}$ is easily achieved
in this configuration. In one series of measurements,
the pump pulse width is controlled with a prism pair compensator
so that the amplitude of the coherent phonon is varied
without changing the fluence. The signal is averaged by a rapid scan system
with an optical shaker typically operating at 20 Hz.

\begin{figure}[bbb]
\centering
\includegraphics[width=7.3cm]{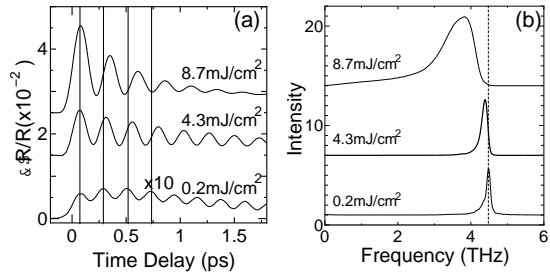}
\caption{ \label{Signal} 
(a) Time-resolved reflectivity changes at three pump fluences 0.2, 4.3, and
 8.7 mJ/cm$^{2}$. 
(b) Corresponding Fourier transformed power spectra.}
\end{figure}

 Figure \ref{Signal}(a) shows the time-resolved reflectivity changes
at different pump power densities.
The period of phonon oscillation is prolonged
and its decay time is shortened with fluence.
Vertical lines are located at the peak positions of the coherent phonon signal
for the lowest fluence,
and the change in peak positions with the oscillation period
is distinctly observed.
The corresponding Fourier transformed (FT) spectra are shown in
Fig. \ref{Signal}(b). 
The peak frequency of the A$_{\rm 1g} $ mode down-shifts and the spectrum
shape is asymmetrically broadened to the lower frequency side. 

\begin{figure}[tb]
\includegraphics[width=7.3cm]{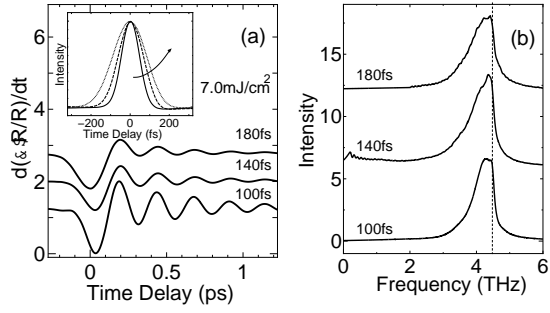}
\caption{\label{Amplitude} 
Dependence of time derivative of reflectivity change on pump pulse width. 
(a) Temporal profiles for three pump pulse widths.
Inset shows cross-correlation trace of pump and probe pulses.
(b) Corresponding Fourier power spectra.  
}
\end{figure}

 The effect of anharmonicity of the lattice potential on softening
of the A$_{\rm 1g} $ mode is examined by controlling the phonon amplitude.
The width of the pump pulse is adjusted by a prism pair, while the width
of the probe pulse is maintained at 100fs.
As the pump pulse width increases with the period of phonon oscillation,
the amplitude of the coherent phonon drastically decreases. 
Figure \ref{Amplitude}(a) shows the coherent phonon signal for three pump
pulse widths at a fluence of 7.0 mJ/cm$^{2}$. The pump pulse width is
evaluated by the width of the cross-correlation trace with probe-pulse and
the time traces are labeled with these values.
The corresponding cross-correlation traces are displayed in the inset. 
The amplitude of the phonon oscillation decreases by several times
with the pump pulse width; however, the temporal profiles show negligible change.
Figure \ref{Amplitude}(b) shows the corresponding Fourier power spectra obtained
after removing the monotonic component. We also verified the dependence on
the pump pulse width at higher fluences up to 17.3 mJ/cm$^{2}$,
which is immediately below the damage threshold,
and found that the temporal profile shows no significant change.
It is clearly shown that the softening of coherent phonons is mostly determined
by the accumulation of the pump pulse energy and not by the anharmonicity
of the lattice potential.

 To analyze the time-dependent frequency shift, we numerically fit the data
over a monocycle of phonon oscillation with a function expressed
as the sum of damping oscillation
and a single exponential decay, and determine the instantaneous frequency.
The region of the monocycle is selected approximately by the naked eye,
and the frequency thus obtained is insensitive to the detail of the selection.

\begin{figure}
\includegraphics[width=6.0cm]{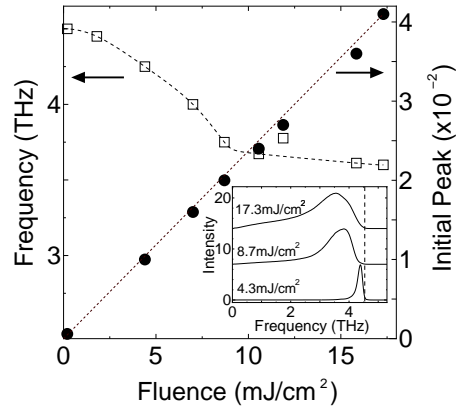}
\caption{\label{Plot} 
Fluence dependence of initial frequency of oscillation (open squares)
and height of initial peak (filled circles).
Inset shows the Fourier power spectra at three pump fluences
of 4.3, 8.7, and 17.3 mJ/cm$^{2}$.
}
\end{figure}

 Open squares shown in Fig. \ref{Plot} indicate the fluence dependence
of initial frequency of the coherent phonon obtained by monocycle fitting.
The frequency superlinearity decreases with fluence and shows saturation
above 10 mJ/cm$^2$.
Excitation exceeding 20 mJ/cm$^2$ leads to optical damage of the sample.
Filled circles in Fig. \ref{Plot} indicate the fluence dependence of
the height of the initial peak.
The figure shows that the initial peak height increases linearly with fluence,
which indicates that there is no significant absorption saturation.
Therefore, the saturation of frequency shift originates
from a purely internal mechanism.
Fourier power spectra at the corresponding fluence is displayed
in the inset of Fig. \ref{Plot}.
The spectrum shows a large frequency shift and broadening
as the fluence increases from 4.3 to 8.7 mJ/cm$^2$.
Compared with the spectrum at 8.7 mJ/cm$^2$, the frequency shift
of the spectrum at 17.3 mJ/cm$^2$ is saturated,
although remarkable growth of the low frequency component materializes.

\begin{figure}[b]
\includegraphics[width=5.8cm]{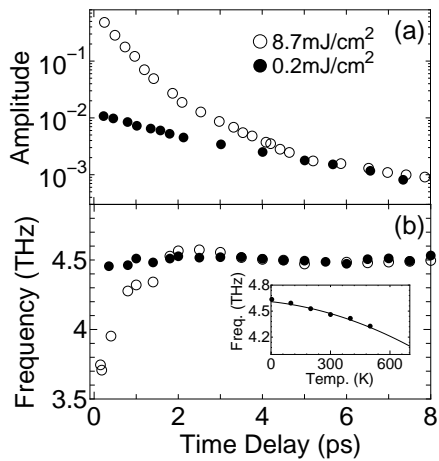}
\caption{\label{t-Amplitude} 
Time evolutions of amplitude (a) and frequency (b) of
coherent A$_{\rm 1g}$ phonon for two fluences.
The inset represents the temperature dependence of A$_{\rm 1g}$ phonon frequency.
Filled circles and open circles correspond to pump fluences
of 0.2 and 8.7 mJ/cm$^{2}$, respectively.
}
\end{figure}

 Figure \ref{t-Amplitude} shows time evolutions of the amplitude
and frequency of the coherent A$_{\rm 1g}$ phonon for two fluences.
Although the amplitude decreases almost exponentially
with a decay constant of 0.4 ps$^{-1}$ at 0.2 mJ/cm$^{2}$,
it shows an accelerated decay for the first 6 ps at 8.7 mJ/cm$^{2}$,
as shown in Fig. \ref{t-Amplitude}(a).
The time dependence of frequency at 8.7 mJ/cm$^{2}$ indicates an abrupt change
and a slight overshoot of the phonon frequency around the time delay of 2 ps
in contrast to the flat time evolution at 0.2 mJ/cm$^{2}$,
as shown in Fig \ref{t-Amplitude}(b).
At higher fluences, an abrupt change in frequency also appears
around the time delay of 2 ps
and becomes more discontinuous. After these transient responses,
the frequency reaches a thermal equilibrium value of 4.50 THz,
which lasts onward and the corresponding temperature is very close
to room temperature.
These experimental results, the saturation behavior of the softening
and the unusual time evolution of phonon frequency, indicate
that the mechanism of the softening process
involves structural change rather than carrier-phonon interaction.

\begin{figure}[b]
\includegraphics[width=4.4cm]{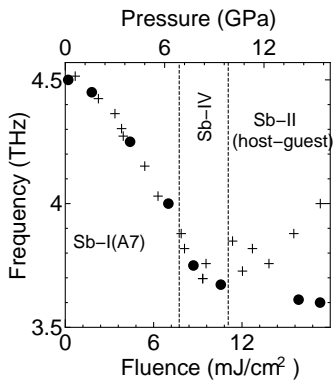}
\caption{\label{power and pressure} 
Comparison of fluence and pressure dependences of phonon frequency.
Crosses show pressure dependence of Raman frequencies for Sb.
Filled circles show pump fluence dependence of
coherent phonon frequencies for Sb.
}
\end{figure}
 
 The change in the phonon frequency of the semi-metal that is initiated
by the structural change
under a high-pressure condition was extensively studied by Raman scattering
and pump-probe experiments \cite{Wang06, Degtyareva07, Kasami06}.
Figure 5 shows the comparison of the fluence dependence of
the initial frequency of the coherent phonon
with the uniaxial pressure dependence of the static phonon frequency measured
by Raman scattering \cite{Degtyareva07}. Crosses in the figure indicate
the pressure dependence of the Raman peak frequency,
and the pressure is indicated in the upper horizontal axis.
Vertical dashed lines located at the pressures of 7.2 and 9.0 GPa
correspond to the phase boundaries
of Sb-I (hexagonal setting), Sb-IV, and Sb-II (host-guest structure).
Filled circles indicate the pump fluence dependence of the phonon frequency,
and the fluence is indicated in the lower horizontal axis.
We set the conversion ratio of fluence to pressure
as 0.9 GPa per 1.0 mJ/cm$^{2}$ so that the two curves overlap.
Surprisingly, not only the decreases in super linearity at the Sb-I phase
but also the saturation behavior at phase IV show extremely good agreement
with the fluence dependence.
This correspondence between the transient frequency
and static pressure shifts of the phonon frequency clearly indicates
that the optical high density excitation
generates a compressed state whose pressure is proportional to the fluence.
Since there is no absorption saturation, as shown in Fig. \ref{Plot},
the disagreement of the frequency of the coherent phonon corresponding
to the Sb-II phase could be explained
by the formation of a quasi-stable state instead of a host-guest structure.

 In contrast to the generally accepted idea that high density excitation
weakens the covalent bond and the lattice undergoes
the same expansion process as it undergoes when subject to a heating condition,
time-resolved electron diffraction measurements have revealed
instantaneous contraction
of the atomic distance in graphite \cite{Carbone08, Raman08}.
Theoretically, the contraction in graphite \cite{Jeschke01}
and the mechanism is now considered as the efficient
LA phonon emission by the double resonant
Raman process \cite{Carbone11, Saito02}.
A similar mechanism is also expected in our case; the lattice
contraction explains the good correspondence between
the fluence and pressure dependences of the phonon frequency,
as shown in Fig. \ref{power and pressure}.

 From the result of the optically generated high-pressure state,
the following consideration is provided
for the energy balance and the surface dynamics.
The conventional material parameters used below are taken
from the literature \cite{Landolt}.
The sheet energy density of the strain is roughly
expressed as $u$ = $p^{2}d$/2$C\rm_{33}$,
where $p$ is the pressure, $d$ is the penetration depth,
and $C\rm_{33}$ is the elastic modulus along the c-axis.
Here, we calculate the strain energy density
at the excitation density of 10 mJ/cm$^{2}$.
Using the pressure $p$ of 9.0 GPa, the elastic modulus $C\rm_{33}$ of 44.6 GPa,
and the optical penetration depth of 13 nm,
the resultant strain energy density is 1.2 mJ/cm$^{2}$.
$p$ is deduced from the frequency shift
shown in Fig. 5; the elastic modulus $C\rm_{33}$ is obtained
from the ultrasonic measurement; and the optical penetration depth
is defined as the inverse of
the absorption coefficient ($1/\alpha$) of the amorphous film
of antimony at a wavelength of 790 nm.
Since the photoabsorbed energy is determined as 3.0 mJ/cm$^{2}$
from the optical reflectivity of 70 \%,
it is concluded that a large amount of the absorbed energy is converted
into strain energy for the lattice,
which is consistent with the X-ray experimental result \cite{Reis01}.
It is well known that a strain pulse with a picosecond duration
can be generated by irradiating a metal film
or a semiconductor with intense femtosecond laser pulses
and that it propagates into the sample \cite{Baumberg97, Kasami04, Akimov06}.
Because our experimental conditions are very close to this situation,
a strain pulse is also expected,
and the lattice contraction corresponds to the initial stage of the strain pulse.
The spatial size of the strain pulse estimated
by the optical penetration depth $d$ is 13 nm,
and the acoustic velocity $v$ of antimony along the c-axis is about 2584 m/s.
Therefore, the escape time of the strain pulse
from the optical penetration layer is calculated as 5.0 ps.
The complete recovery from phonon softening and the small temperature
increase observed after 5 ps,
as shown in Fig. \ref{t-Amplitude}, are satisfactorily explained
by the strain pulse,
which carries away the structural deformation
and the associated energy from the surface.
Further theoretical and experimental research is needed to clarify
the dynamics of highly photoexcited solid surfaces.
It is strongly emphasized that the high-pressure
Raman scattering experiment is very powerful
for understanding high density excitation phenomena.

 In conclusion, we have investigated the impulsive softening process
in antimony by using a precise measurement system based on
rapid scan and microscope optics.
The detailed analysis of the instantaneous frequency shift yields
unexpected findings.
The fluence dependence of the initial frequency
is strongly correlated with the high-pressure Raman scattering experiment
and the time evolution
of the phonon frequency shows an abrupt change and overshoot.
These results clearly indicate that the highly excited surface
layer undergoes contraction
and successive expansion processes that leads to ablation.
Transient high pressure associated with lattice contraction represents
the extreme reduction in the phonon frequency and the time evolution
of the phonon frequency indicates the lattice deformation dynamics.
Simple estimation of the energy balance reveals that most of
the photo-injected energy is converted to lattice strain energy
and the duration of the softening process is consistent
with the exit time of the strain pulse propagating deep into the sample.

 The authors thank Professor Takashi Kozasa and Dr. T. Yanagisawa
for stimulating discussions.

\bibliography{sbz}

\end{document}